11.01.07

# First Tests of Thick GEMs with Electrodes Made of a Resistive Kapton

R. Oliveira[1], V. Peskov[1], F. Pietropaolo[1,2], P. Picchi[1,3]
[1] CERN, Geneva, Switzerland
[2] INFN Padova, Padova, Italy
[3] INFN Fracati, Frascati, Itali

## Abstract

We have developed a new design of a GEM-like detector with single-layer electrodes made of a resistive kapton. This detector can operate at gains close to $10^5$ even in pure Ar and Ne and if transited to discharges at higher gains they, due to the high resistivity of electrodes, do not damage either the detector or the front-end electronics. Gains ~ $10^6$ can be achieved in a cascaded mode of the operation. The detector can operate without gain degradation at counting rates up to $10^4$Hz/cm$^2$ and thus it could be very useful in many applications which require safe high gain operation, for example in RICH, TPCs, calorimetric.

## 1. Introduction

Recently developed hole- type gaseous detectors [1-6] open new possibilities in the detection of photons and particles, for example they can operate at relatively high gains in poorly quenches gas mixtures and gases (see for example [4, 5, 7]). The most popular hole- type detector today is the so called Gas Electron Multiplier (GEM) suggested by Sauli [3]. Cascaded GEM structures are now accepted for several large scales high energy physics experiments [8].

In spite of a great progress in the development and optimization of the GEM it is still a rather fragile detector, for example it requires very clean and dust free conditions during its manufacturing and assembling process and could be easily damaged by sparks which are almost unavoidable at high gain operations. Studies performed earlier by us indicated that the maximum achievable gain of the hole- type detectors increased with their thickness [9]. Based on these studies we have developed so a called a “thick GEM” (TGEM) [5, 6, 10]: a metallized from both sides printed circuit board (thickness of 0.5-1.5 mm) with drilled -through holes. This detector allows one to achieve the maximum gain of almost 10 times higher than with the conventional GEM [6]. Later we modified this detector by drilling out a Cu layer around each hole; this allowed one to additionally increase the maximum achievable gain by a factor of ~5. A systematic study of this device and its further improvements were performed by Breskin’s group: instead of drilling out the Cu around the edges of the holes, they manufactured the protective dielectric rims by a lithographic technology [11]. Recently we have developed and tested a TGEM with electrodes coated by a thick layer of graphite paint [12]. We named this detector a Resistive Electrode Thick GEM or RETGEM. The RETGEMT could operate at gains of ~$10^5$; at higher gains it may transit to a streamer mode, however in contrast to violent sparks in conventional GEMs these streamers are mild and did not damage either the detector or the front-end electronics. Certainly there is nothing special in graphite coating and many other resistive materials could be used to achieve the same protective effect. The most important in manufacturing such types of detectors is to use a lithographic technology which ensures high quality and reproducibility of resistive coatings in large scale of productions. In the stream of such efforts we have developed RETGEMs in which Cu or Cr electrodes were coated by CuO or CrO resistive layers [13]. Due to the small thickness, this coating did not provide full spark protection*, however the detector gained more robustness compared to the TGEM.

Both of these RETGEM designs had double- layer electrodes structures: a thin Cu layer and a resistive layer on its top.

The aim of this work is to build and test first prototypes of RETGEMs manufactured by the lithographic technology with electrodes made of single- layer resistive materials.

## 2. Detector's Design and Experimental Set Up

The detectors studied in this work were manufactured from standard printed circuit boards (PCBs) having a thickness of 1, 1.6 or 2.4 mm. On the both surfaces of the PCB sheets of resistive kapton 100XC10E5 50μm thick were glued (the glue FR4). The surface resistivity of this material depending on a particular sample may vary from 500 to 800 kΩ/□. The holes were drilled by a CNC machine as was done earlier in the case of TGEM; they were 0.8 mm in diameter, the pitch was 1.2 mm and the active area of the detector was 30x30 $mm^2$. A Cu frame was manufactured by a photolithographic technique in the surrounding area of the detector in order to provide good electrical contacts with the HV and signal cables-see Fig.1.

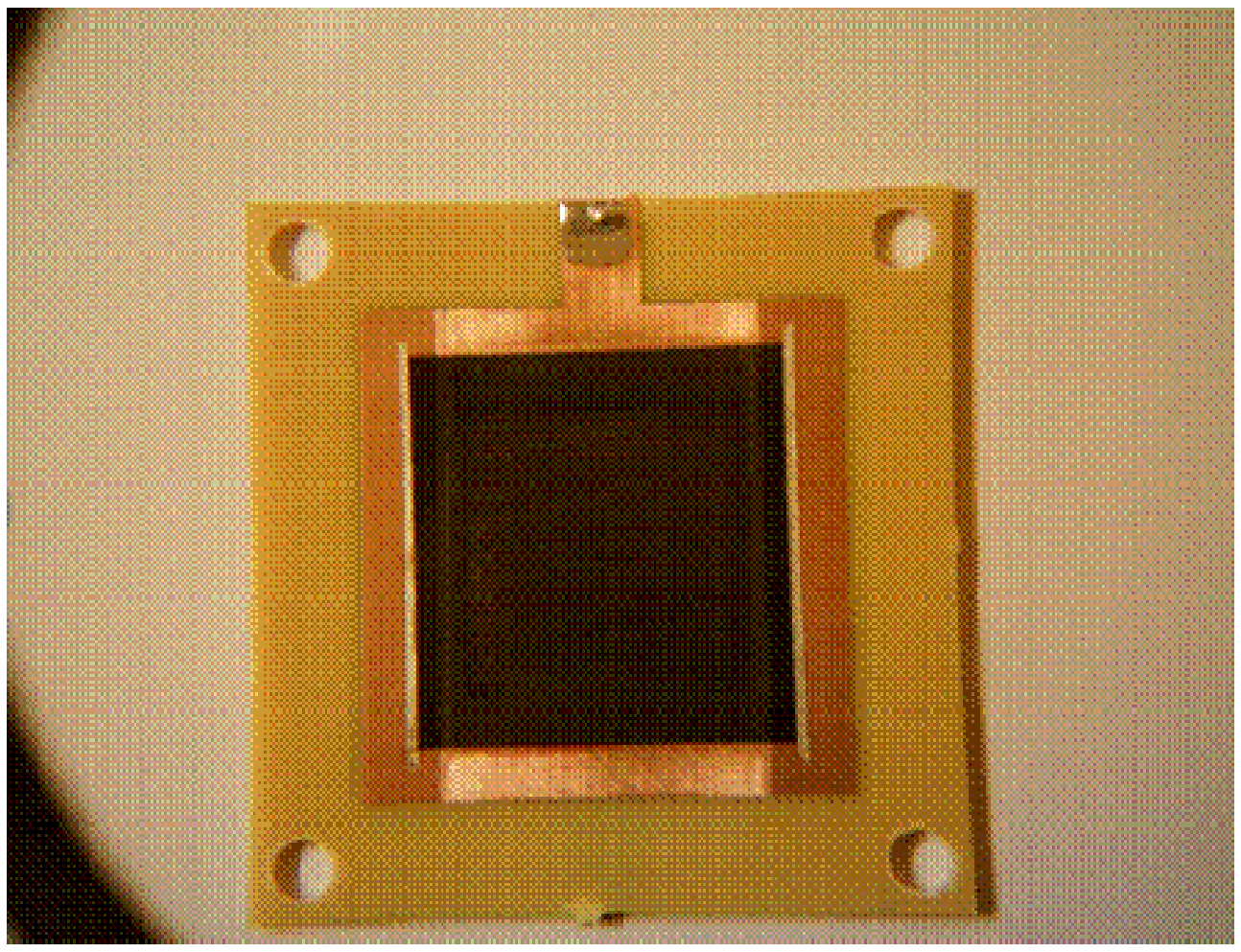

Fig.1. Photo of RETGEM made of a single-layer resistive kapton

* The efficiency of the spark quenching depends on the amount of surface charge from the incoming avalanche that is needed to substantially modify the field in the gap quenching the discharge, or in other words, the resistive layer capacity per unit area. This will be high if the layer is thin or if the dielectric constant is high.

Unfortunately, not all detectors manufactured by this simple technology (we called them" first prototypes") were perfect: some holes contained micropartcles and microwires of caption remaining after the drilling. To avoid these defects we slightly changed the manufactured technology: prior to drilling we glued on the top of the kapton sheets a Cu foil 35 μm thick- see Fig. 2. After the drilling process was finished the Cu foils were etched in the active area of the detector and only a Cu frame for the connection of the HV wires was preserved in the periferical part of the detector –see Fig. 2. This modification allows one to achieve an excellent quality of all the holes. We called the detectors manufactured by this technology "improved prototypes."

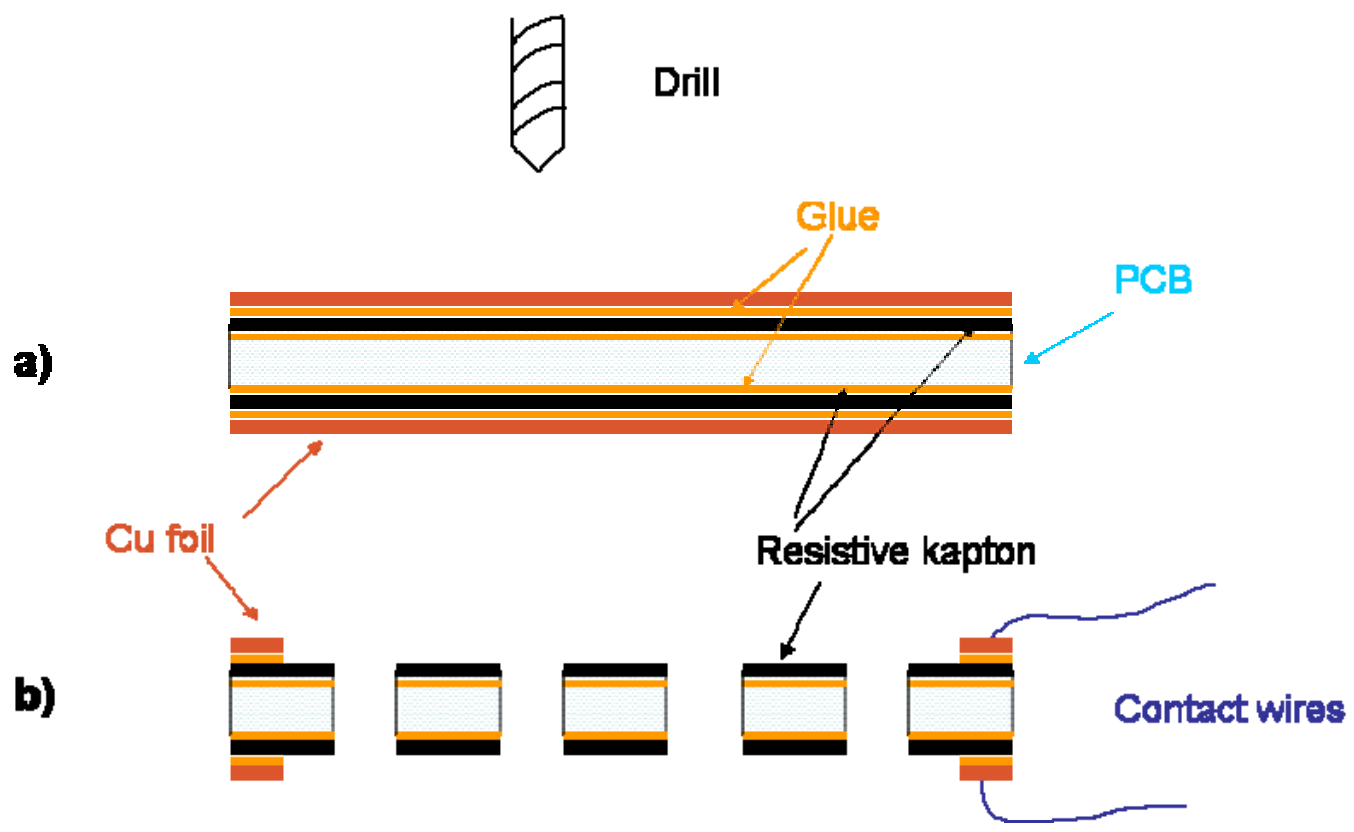

Fig. 2. Two step process in drilling holes in improved RETGEM:
a) On the top of the PCP a resistive caption sheet and a Cu foil were glued; this structure was drilled through by the CNC machine
b) The Cu foil was removed from the active area of the detector

The experimental set up for the testing of the detectors is shown in Fig.3.

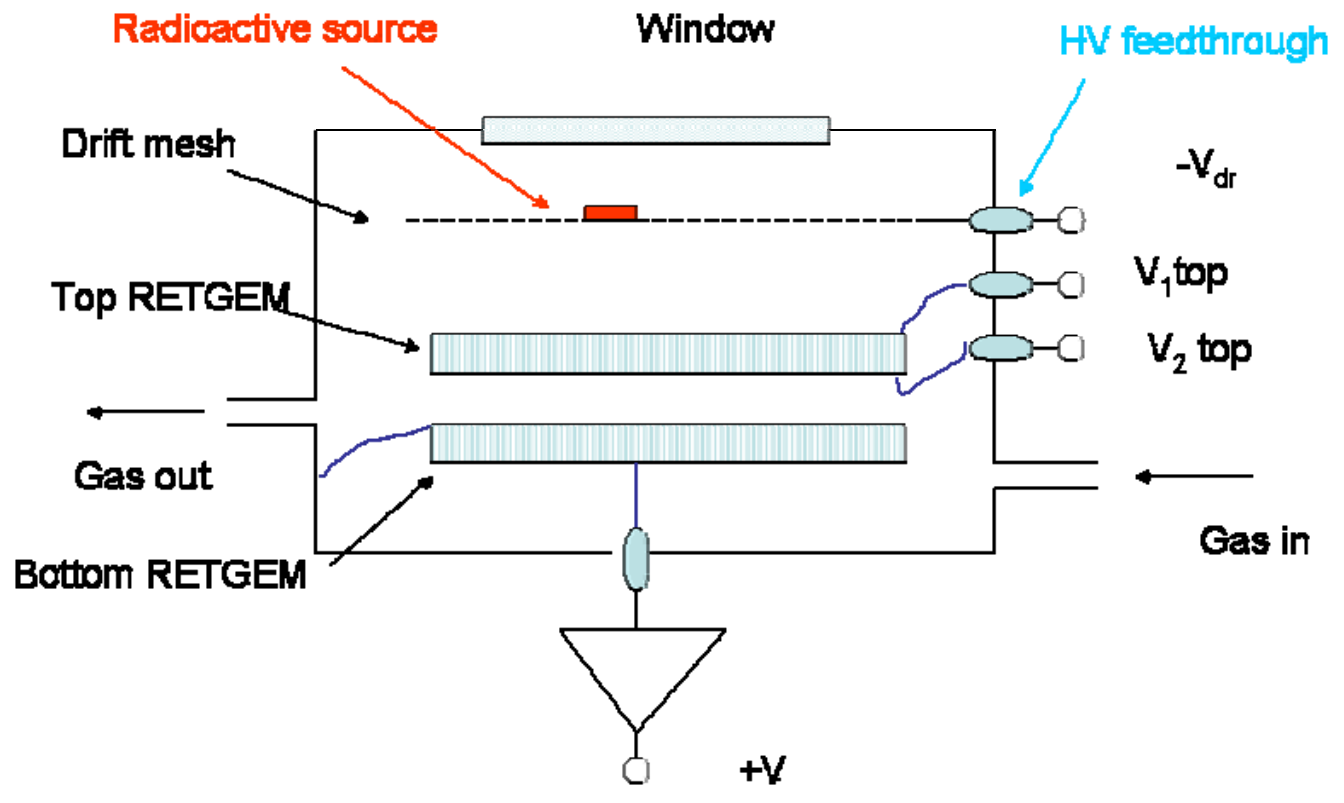

Fig.3. A schematic drawing of the experimental set up

It consists of a gas chamber inside which one or two RETGEMs operating in cascade mode were installed and a gas system allowing the flushing though the chamber of various gases. Most of the tests were performed in Ar, Ar+20%$CO_2$ or in Ne at pressure of 1atm. Ionization of the gas was produces either by a $^{241}$ Am alpha source or by a $^{55}$Fe X-ray source. The signals from the detector were recorded by a charge sensitive amplifier CANBERRA and then, if necessary, additionally amplified by a research amplifier.

## 3. Results

### *3.1. Results Obtained with First Prototypes*

Fig. 4 shows gain vs. voltage measured in the case of the first prototype 1mm thick operating in pure Ne. Measurements were performed using $^{241}$Am at low gains and $^{55}$Fe at high gains. One can see that in the case of the $^{55}$Fe source gains close to $10^5$ were achieved. At higher gains the detector transited to mild streamers which did not harm either the detector or the preamplifier. In Ar which requires much higher voltages this detector at a gain of ~$10^4$ transited to a continuous discharge and the visual observation shows that the discharge was caused by the kapton microwires stick out from one of the imperfect holes. However, this continuous discharge was not harmful as well. In some cases the discharge triggered by the debris inside the holes could propagate along the surface toward the surrounding Cu frame (surface streamers).

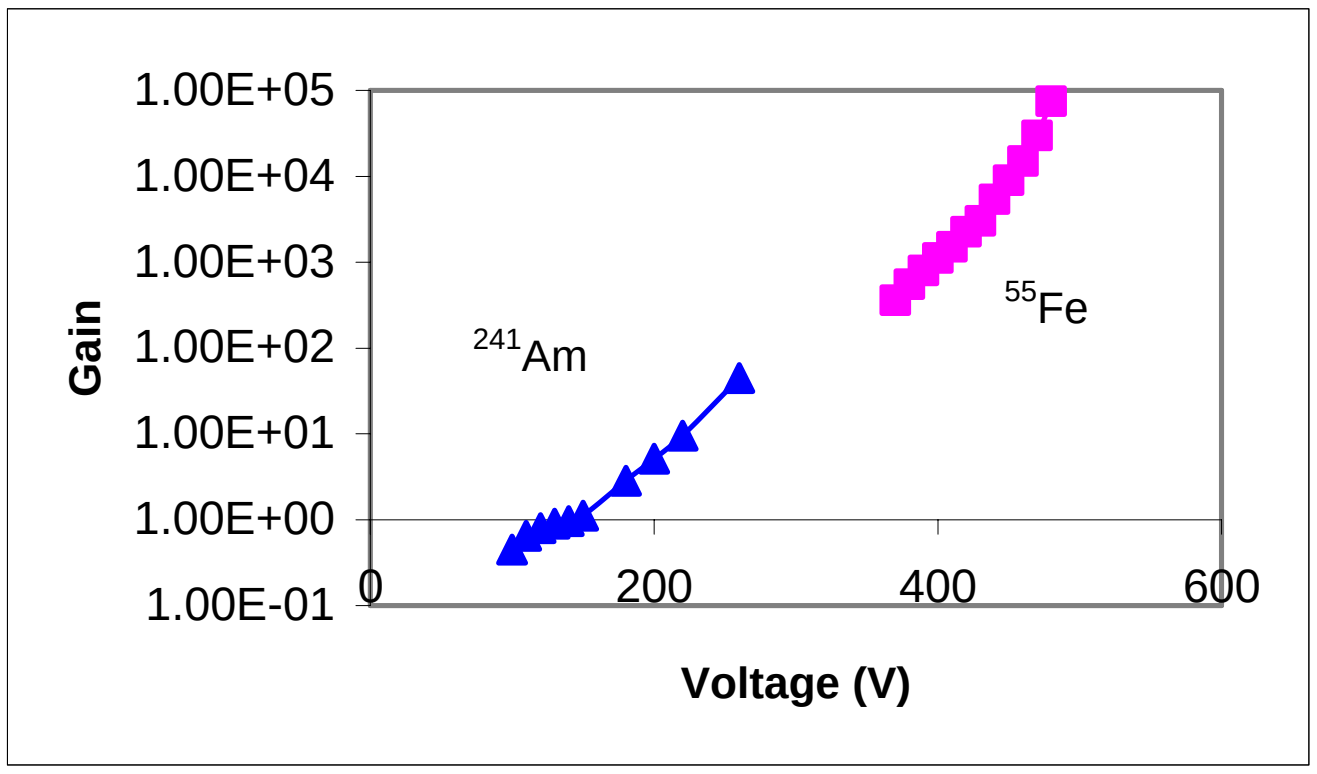

Fig. 4. Gain vs. voltage measured with the first prototype of the RETGEM 1mm thick using a $^{241}$Am and a $^{55}$Fe radioactive sources

Figs.5 and 6 show the results of gain measurements performed with the 2.4 mm thick RETGEM flushed with Ne, Ar or Ar+20%$CO_2$. One can see that a gain close to $10^5$ was achieved in Ne and ~$10^4$ in Ar and in Ar+20%$CO_2$.

It higher gains streamer type of discharges appeared which may transit to a continuous discharge with a further increase of the voltage. None of these discharges were distractive.

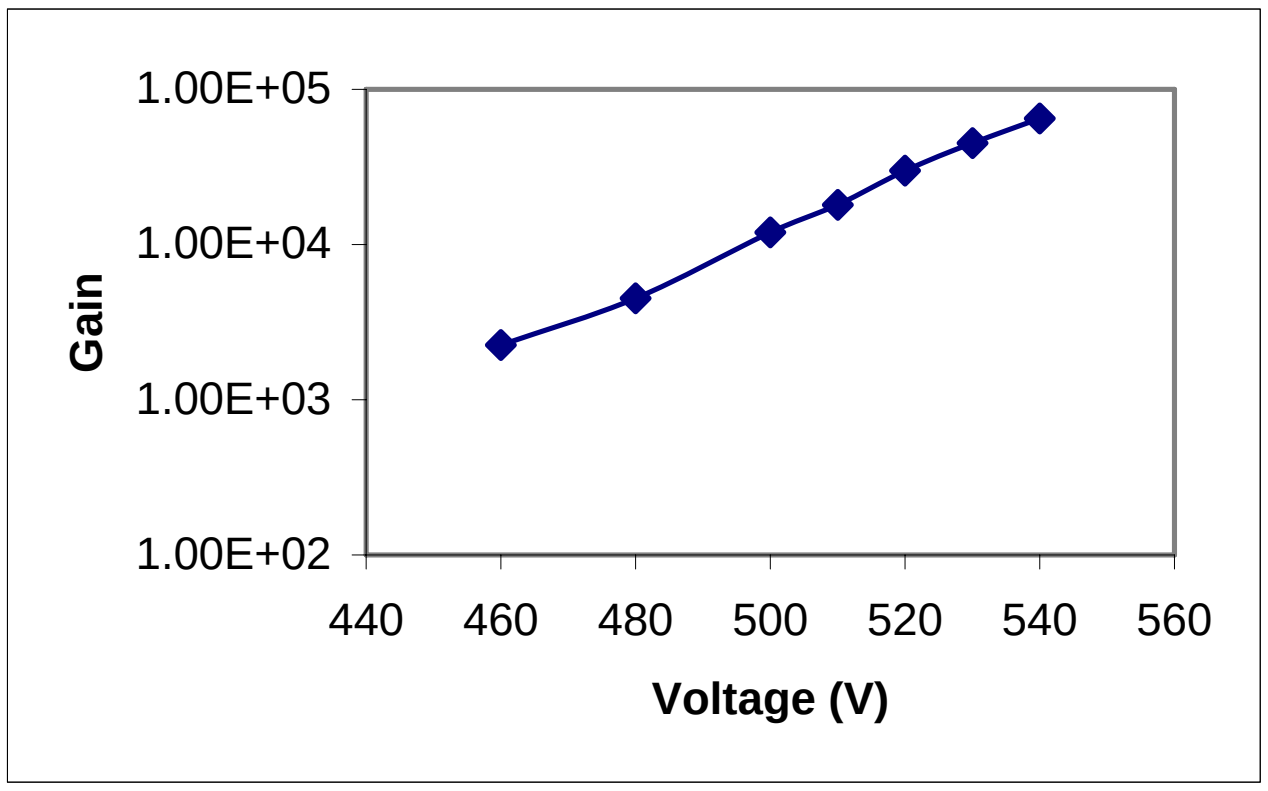

Fig. 5. Gain vs. voltage measured in Ne with the RETGEM 2.4 mm thick. As a radioactive source $^{55}$Fe was used

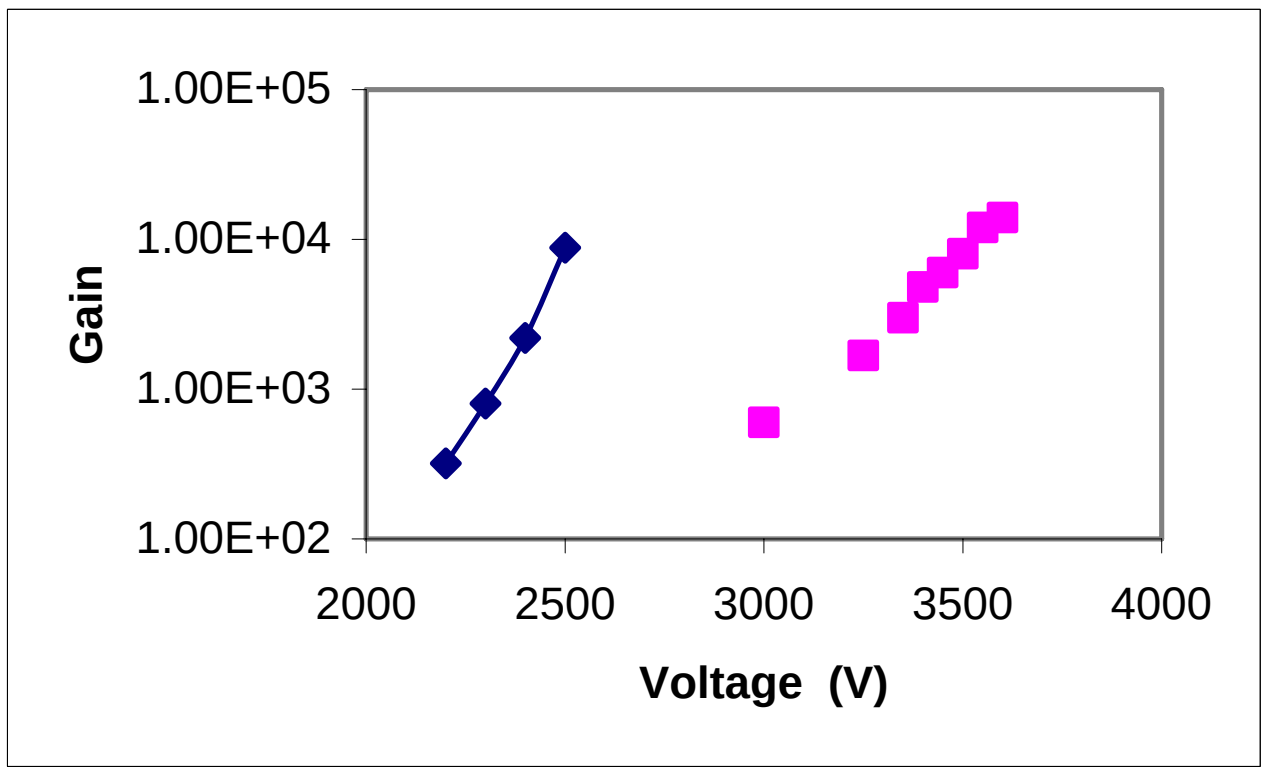

Fig. 6. Gain vs. voltage measured in Ar (blue symbols) and Ar+20%$CO_2$ (rose symbols) with the RETGEM 2.4 mm thick. As a radioactive source $^{55}$Fe was used

### *3.2. Results with Improved Prototypes*

The improved designs of RETGEM allowed us to achieve gains 5-7 times higher than with the first prototypes (see Fig. 7 and 8). At gains close to and above $10^5$, due to the strong space charge effect in the avalanche development, the detector began losing its prortionality and, as in the Geiger counters, the pulses from the detector had a tendency to become equal in amplitudes (see Fig. 9 and 10). A percentage of the pulses began to be accompanied by their successors (see Fig. 11 and 12). If discharges appeared at higher gains they were mostly confined inside the holes. However, if one continues to increase the voltage applied to the detector some of them may transit to surface streamers propagated towards the Cu frame.

Fig. 13 shows signal amplitude as a function of the counting rate measures in Ar. No surface charging up effect was observed up to counting rated of $10^4$Hz/cm$^2$ - the maximum counting rate available from our $^{55}$Fe source.

In the next series of experiments we tested double RETGEM operating in the cascaded mode (see Fig. 2). Some results are presented in Fig.14. In this figure the overall gains is plotted as a function of the positive voltage applied to the bottom RETGEM for three

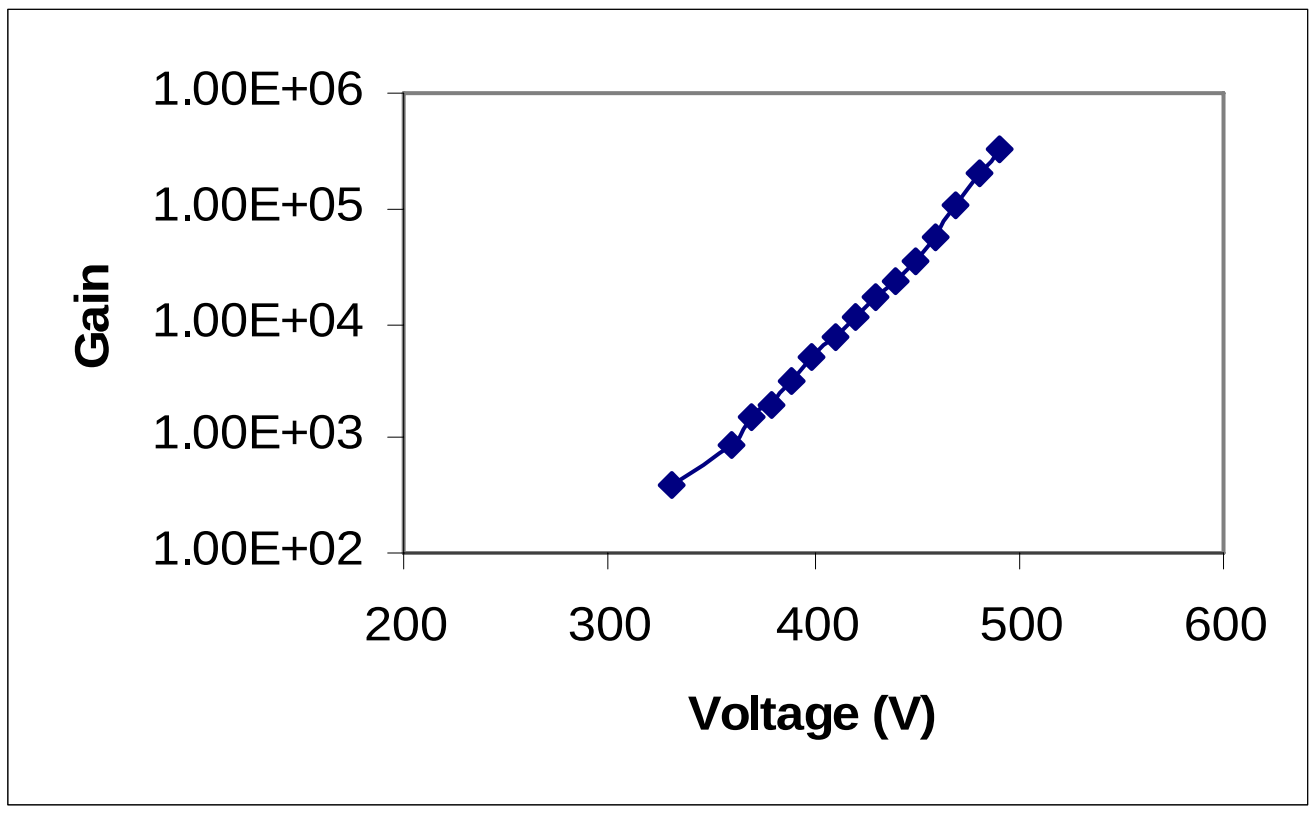

Fig 7. Gain vs. voltage measured in Ne with the improved RETGEM 1 mm thick. Measurements were performed with $^{55}Fe$.

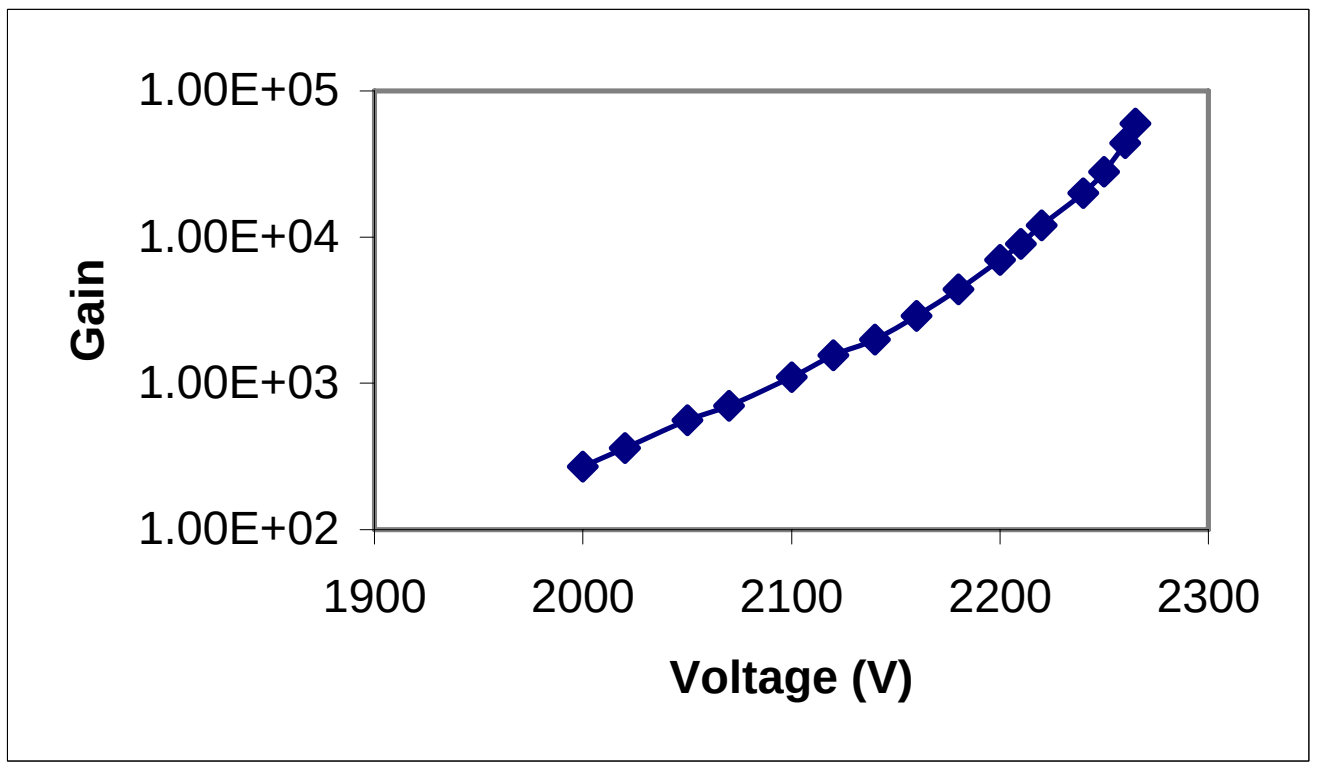

Fig. 8. Gain vs. voltage measured in Ar with the improved RETGEM 1 mm thick. Measurements were performed with $^{55}Fe$.

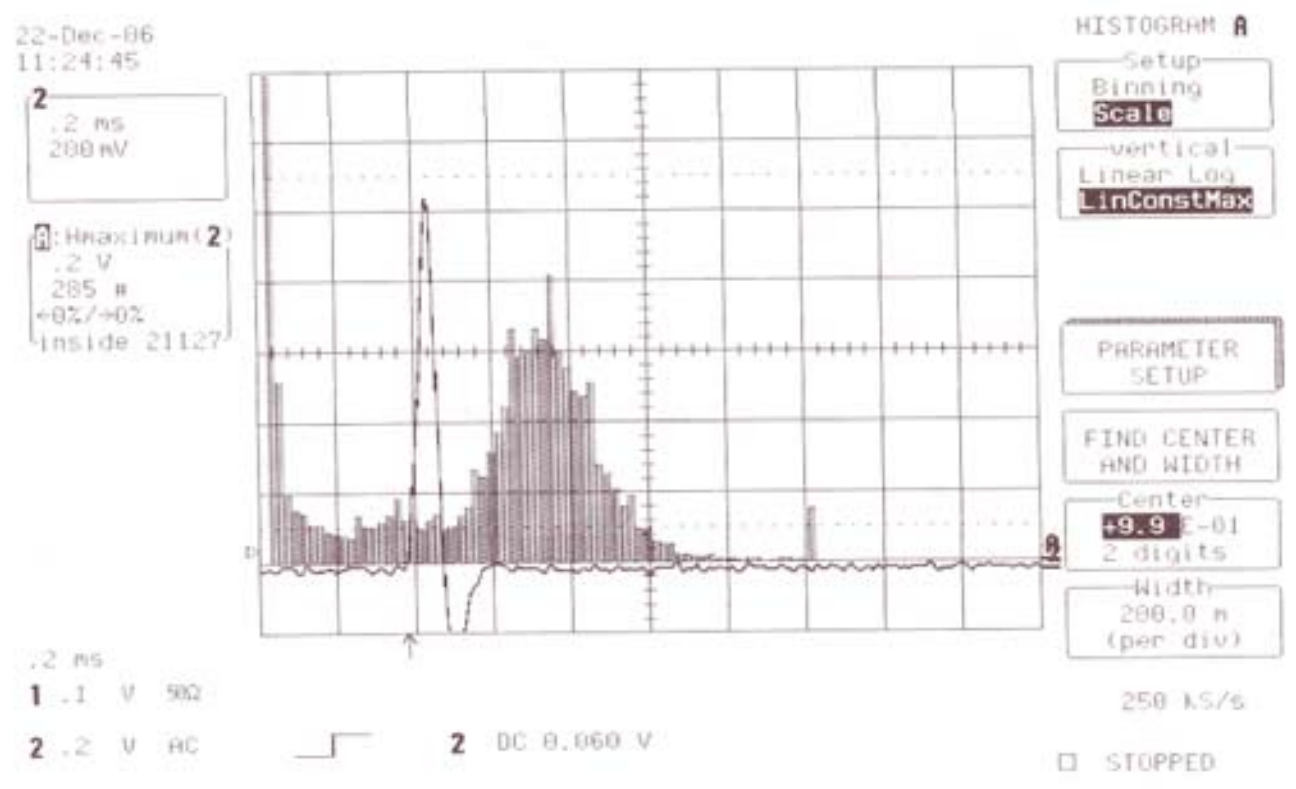

Fig. 9. A pulse height spectrum measured with improved RETGEM operating in Ar at a gain of 1320. All detector area was exposed to the $^{55}Fe$ radiation.

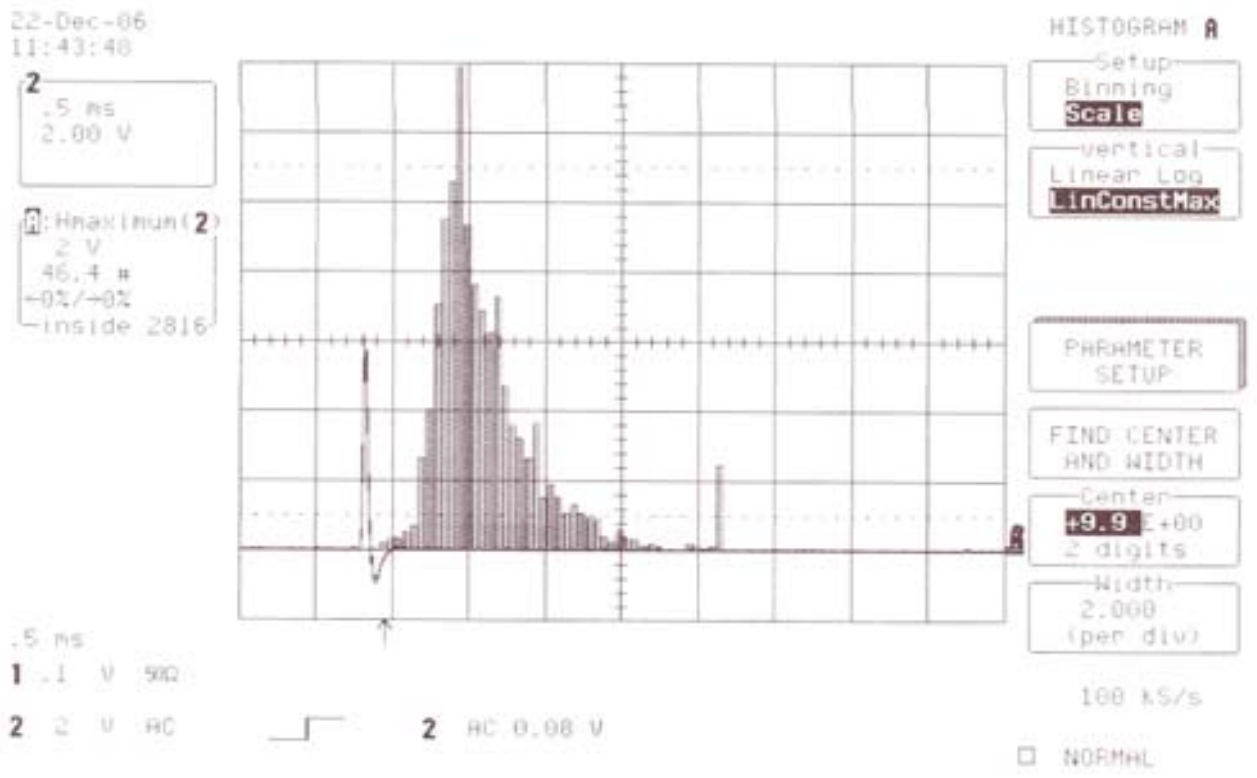

Fig. 10. A pulse height spectrum measured at a gain~$10^5$.

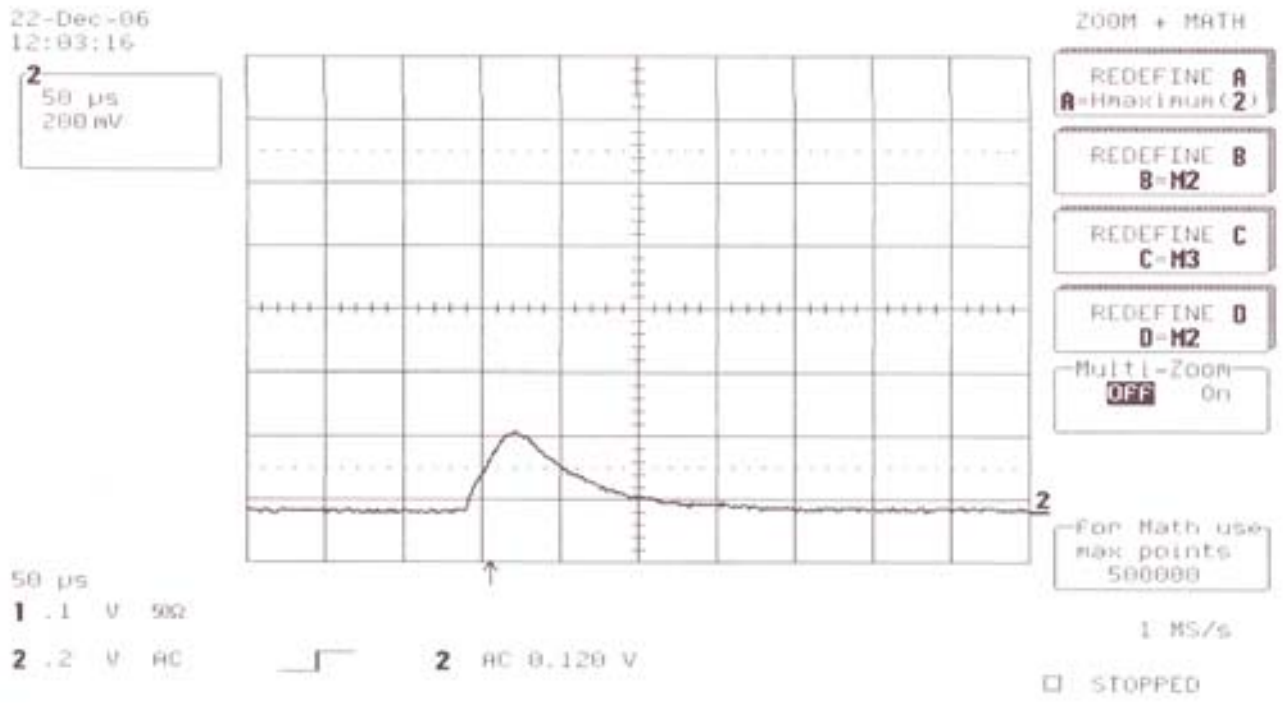

Fig 11. Typical oscillogramm of the signal measured directly from the charge sensitive preamplifier at a gain of 4000.

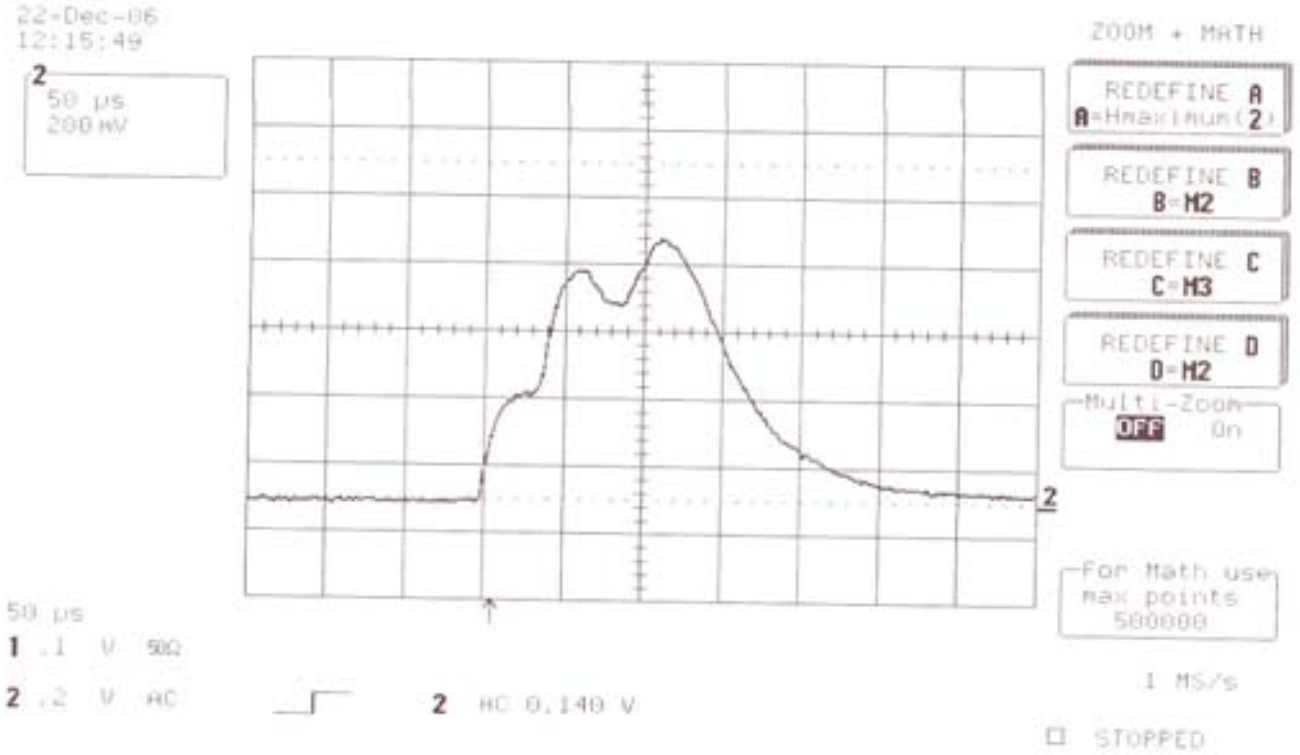

Fig 12. Shape of some pulses appearing at gains > 40000.

different voltages $V_t=V_{1top}-V_{2top}$ (see Fig. 2) applied to the top RETGEM. One can see that gains close to $10^6$ were achieved with the double step RETGEMs. In Ar at higher gains discharges between two RETGEMs may appear triggered by the initial discharge in one of the holes of the bottom RETGEM. As in previous cases, none of these discharges damaged the detector. For example, for several cases we initiated continuous discharges in the holes and between the RETGEM for the total duration of 10 min- see the photo. After the discharge

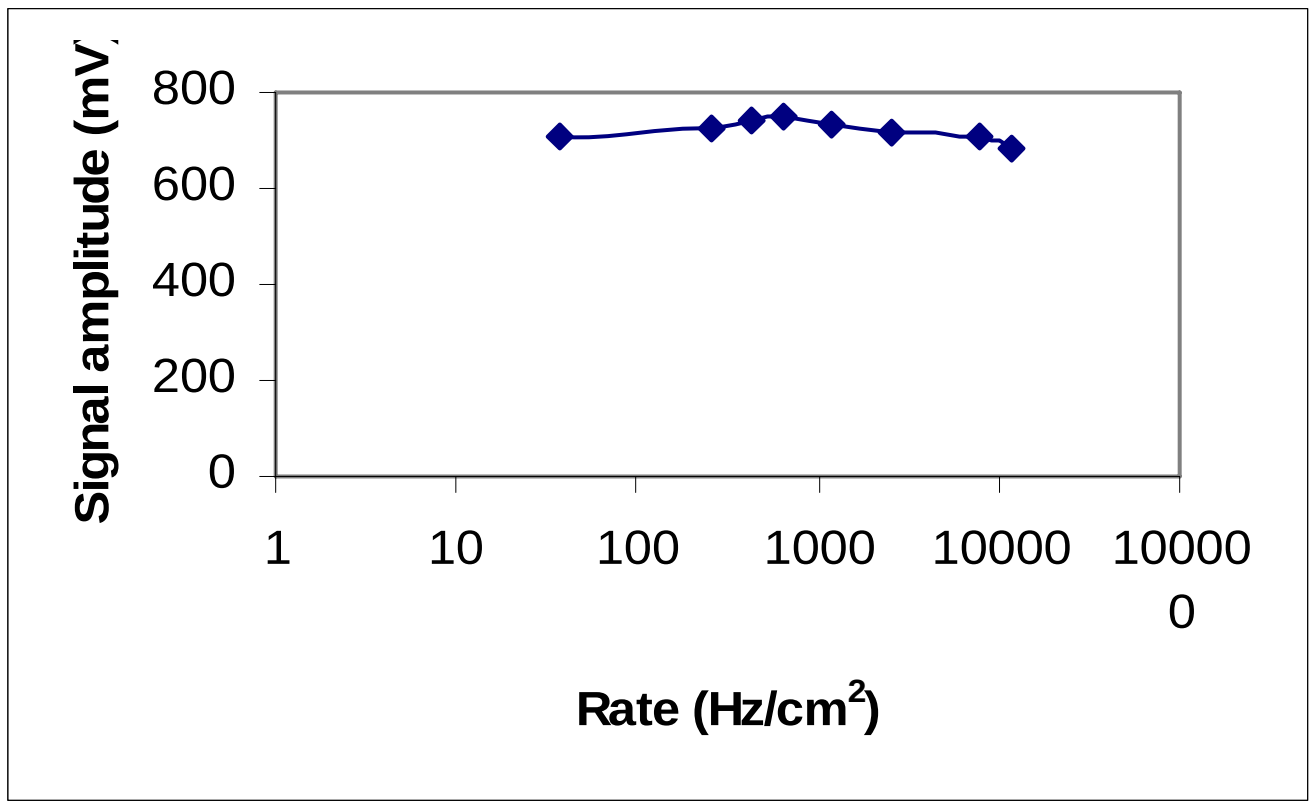

Fig 13. Signal amplitude as a function of the counting rate measured with improved RETGEM operating in Ar

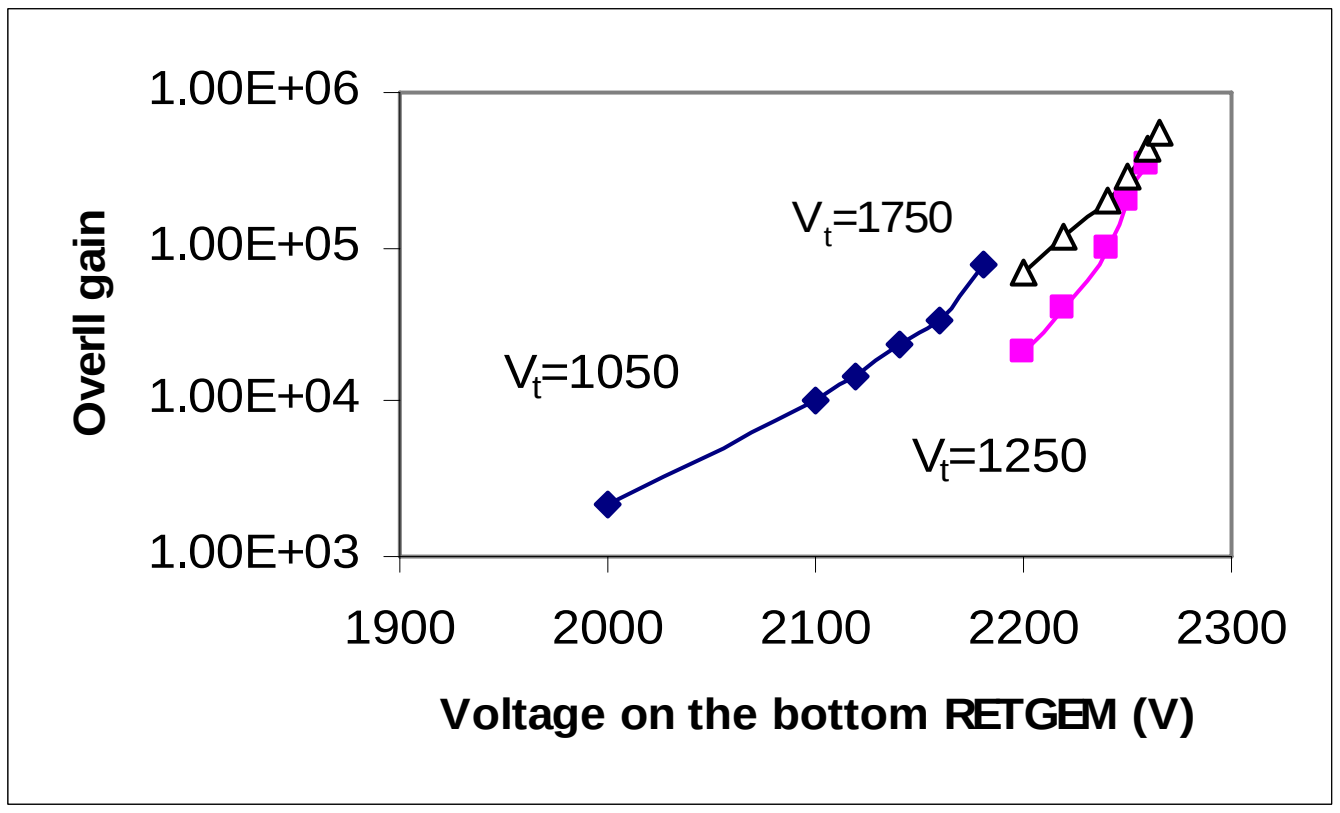

Fig 14. Overall gain of a double RETGEMs operating in Ar at various voltages $V_t$ applied to the top RETGEM

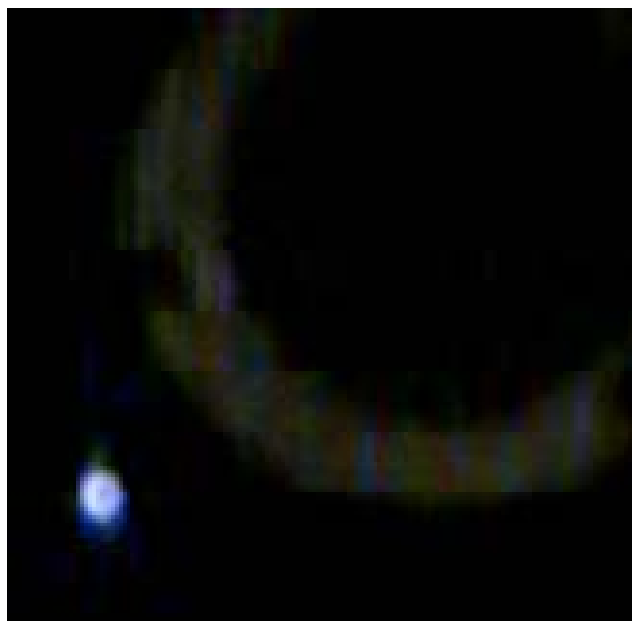

Fig. 15. A photo of a continuous discharge occurring between two RETGEMS at overall gain $>10^6$. After ~10 min of this discharge the detector continue to operate normally

was stopped (by reducing the voltage on the detector's electrodes) the RETGEMs continued to operate without any change of their characteristics including the maximum achievable gain.

## 4. Discussion and Conclusions

For the first time in this work we have developed and successfully tested GEM –like detectors with electrodes made of single -layer resistive material. Measurements show that these detectors allow one to reach only slightly higher gains than with TGEM or with double- layer RETGEMs. This is because in all of these detectors the transition to the discharges occurs when the Raether limit is met at:

$$An_0 \sim 10^8 \text{ electrons,}$$

where A is the gas gain and $n_0$ is the number of primary electrons created in the gas by the radioactive source (see [9] for more details).

However, in contrast to GEMs or TGEMs where sparks may cause fatal destructions, the single- layer RETGEMs were fully spark protected.

An interesting effect was observed during these studies: the discharges created in the holes of the RETGEM may propagate along the kapton surface to the Cu frame. Similar surface streamers were observed by us earlier in the case of microstrip gaseous detectors [14, 15]. Studies show that surface streamers can easily propagate along dielectric surfaces on large distances even in rather weak electric fields [15, 16]. We are now developing another design of the RETGEMS in which the surface steamers will be strongly suppressed.

It was also observed in this work that in the case of the double RETGEMs operating in pure Ar and Ne the discharge in the hole of the bottom RETGEM may trigger discharges between the RETGEMs. Similar effects were observed earlier in the case of double GEMs and this phenomena is well understood today (see for example [17, 18] and references therein). This type of discharge could be avoided by the optimization of the voltages applied to the top and bottom RETGEMs as well as decreasing the voltage between the RETGEMs (or increasing the distance between them) [17].

In conclusion, we would like note that achieved gains $10^5$-$10^6$ are sufficient for most applications. The RETGEM is very robust, can be assembled in dusty conditions, does not require special cleanness of it surfaces or the gas chamber and the gas system and can operate in poorly quenched gases.

Thus we believe that RETGEMs could be very useful in many applications which require safe and high gain operations, for example in RICH, TPCs, calorimetry, etc. Certainly, other resistive coatings could be used as well and the work for their search and study will be the subject of our future projects.